%% file: 1992_camera_ready.tex
\documentclass[a4paper]{article}

\usepackage{INTERSPEECH2020}
\usepackage{amssymb,amsmath,graphicx,bbm,multicol,subcaption}

\title{On Front-end Gain Invariant Modeling for Wake Word Spotting}
\name{Yixin Gao, Noah D. Stein, Chieh-Chi Kao, Yunliang Cai, Ming Sun, Tao Zhang, Shiv Vitaladevuni}
\address{
Alexa Speech, Amazon
\email{\{yixigao,steinns,chiehchi,cyunlian,mingsun,taozhng,shivnaga\}@amazon.com}
}

\begin{document}

\maketitle
\begin{abstract}
Wake word (WW) spotting is challenging in far-field due to the complexities and variations in acoustic conditions and the environmental interference in signal transmission. A suite of carefully designed and optimized audio front-end (AFE) algorithms help mitigate these challenges and provide better quality audio signals to the downstream modules such as WW spotter. Since the WW model is trained with the AFE-processed audio data, its performance is sensitive to AFE variations, such as gain changes. In addition, when deploying to new devices, the WW performance is not guaranteed because the AFE is unknown to the WW model. To address these issues, we propose a novel approach to use a new feature called $\Delta$LFBE to decouple the AFE gain variations from the WW model. We modified the neural network architectures to accommodate the delta computation, with the feature extraction module unchanged. We evaluate our WW models using data collected from real household settings and showed the models with the $\Delta$LFBE is robust to AFE gain changes. Specifically, when AFE gain changes up to $\pm$12dB, the baseline CNN model lost up to relative 19.0\% in false alarm rate or 34.3\% in false reject rate, while the model with $\Delta$LFBE demonstrates no performance loss.
\end{abstract}
\section{Introduction}
\input{introduction}
%
%
\section{Method}
\input{method}

\section{Experiments}
\input{experiments}
%
\section{Conclusions}
In this paper we study the influence of audio front-end (AFE) variation on far-field wake word (WW) spotting and propose a modeling approach for AFE gain invariant WW spotting. For edge devices such as smart speakers,  AFE improves the SNR of the signals from complex acoustic environments and helps to mitigate the challenges in far-field. Since the wake word spotter depends on the AFE, any change in AFE may cause degradation in wake word performance. In this paper we proposed a novel approach called $\Delta$LFBE by implementing new modeling architectures which are equivalent to taking the temporal difference of contiguous input LFBE frames. We showed both in theory and by experiments that the $\Delta$LFBE effectively eliminates the influence by AFE gain changes and stabilizes the performance of WW spotting . The $\Delta$LFBE feature decouples AFE design and WW model building and helps to deploy WW models to devices with different AFEs. Besides WW spotting, the benefits of the $\Delta$LFBE generalize to other tasks such as audio tagging.

The $\Delta$LFBE not only provides invariance to AFE gain changes, but additional Mel-band-wise invariance. Future work includes exploring the effects of $\Delta$LFBE in other types of AFE changes in real use cases. 

\bibliographystyle{IEEEtran}
\bibliography{mybib,wwbib}

\end{document}

%% file: introduction.tex
Wake word (WW) is the gatekeeper that enables users to interact with the cloud-based smart assistants through voice-enabled devices like Amazon Echo, Google Home, Apple HomePod, etc. 
One of the major challenges for those devices to scale to millions of households is to cope up with the unknown acoustic conditions in users' homes, which include varying levels of acoustic echo, noise, reverberation, and interference which can significantly impair the WW detection in spoken utterances.

\begin{figure}[t]
 \vspace{-0.5em}
 \centering
 \includegraphics[width=0.43\textwidth]{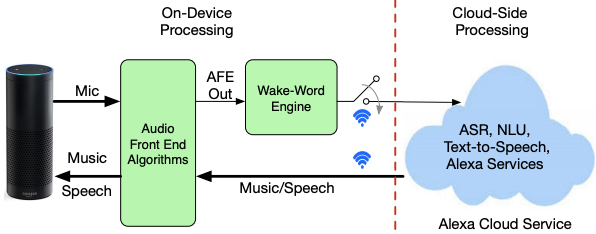}
 \caption{The Alexa voice service system described in \cite{Chhetri2018}.}
 \label{fig:system}
 \vspace{-1.5em}
 \end{figure}

Significant progress has been made in recent years to improve far-field WW detection (also known as keyword spotting), such as investigating novel model architectures \cite{Chen2014_DNN, Sainath2015_CNN, He2017_Seq2Seq, Sun2017_TDNN,Kumatani2017_RAW,Shan2018_Attn,Guo2018_Highway, Wu2018_Monophone}, improving the training efficiency \cite{ Panchi2016_Multitask,Tucker2016_SVD,Raju2018_Augm,Li2018_TS, Chen2018_MAML}, as well as designing the audio front-end (AFE) algorithms to mitigate the challenges under a variety of acoustic conditions. A suite of key algorithms in a multichannel audio front-end were discussed in \cite{Chhetri2018}
for the Alexa voice services systems as shown in Figure \ref{fig:system}, where the AFE output is the input to the WW engine.  Note that the AFE is highly customized to the device, depending on the computational budget. The behavior of the AFE may vary from one device type to another. 
 
Since the WW model ingests the AFE-processed audio and the AFE is hardware dependent,
the WW performance is sensitive to the any change in the AFE. Ideally a jointly optimized AFE and WW for each device would achieve the best performance, however it is neither efficient nor scalable. In practice,  in order to improve the AFE robustness,  the WW model is usually trained and tuned with data collected from a large group of devices with different AFEs. However if there is an AFE update, the WW performance may be affected. Moreover, when we deploy the WW model to a new device, the performance is not guaranteed since the AFE is unknown to the WW model. In fact, we will show in Section \ref{sec:expr} that changing the  AFE gain may cause severe WW performance degradation or operating point (OP) shifting, therefore harming the customer experiences. 

Much work has been done in the literature to improve the robustness to loudness variation of the audio, such as a trainable AFE based on Automatic Gain Control (AGC ) \cite{perez2011automatic,prabhavalkar2015automatic} and Per-Channel Energy Normalization (PCEN) \cite{wang2017trainable,battenberg2017reducing,lostanlen2018per,lostanlen2019long}. PCEN estimates the local magnitude per time step and frequency band using a infinite
impulse response (IIR) filter and has been applied to keyword spotting \cite{wang2017trainable} and ASR \cite{battenberg2017reducing}. The authors in \cite{lehner2018online} showed that some loudness related features like 0-th MFCC will lead to increase of false positive 
and proposed to use a carefully selected set of features for vocal detection in mixed music signals. Another family of approaches such as cepstral mean subtraction (CMS) and cepstral mean-variance normalization (CMVN) \cite{Atal1974cmn,liu1993cmn,Rehr2015CNS,kalinli2019parametric} track the background noise level over a certain period of time like 15 seconds. The effectiveness of these approaches depends on the initial estimate or requires sufficient history to adapt precisely. If a gating mechanism is present, for example a VAD unit, or in a cascade architecture \cite{cascade2017}, CMS is less effective since the available history is limited.

In this paper we address the challenges caused by AFE variance from a wake word modeling perspective and build gain-invariant WW spotters. In our proposed WW models, instead of using the traditional logarithm Mel-filterbank energy (LFBE) features, we use a new feature called $\Delta$LFBE, which is the temporal difference of two contiguous LFBE frames. We show that the $\Delta$LFBE feature is invariant to a constant scaling of audio volume that is typically caused by AFE gain changes. Moreover, the $\Delta$LFBE feature provides additional invariance to filters whose magnitude response varies slowly within each Mel band. We also show that the delta operation can be folded into the computational graph of a neural network 
so that no extra development effort is needed for feature extraction. 
To the best of our knowledge, this is the first attempt to solve AFE gain invariance in wake word spotting. Additional experiment on a public dataset for an audio tagging tasks indicates that the benefit of gain-invariance of the $\Delta$LFBE generalizes to more tasks.

%% file: method.tex
\label{sec:dlfbe}
\subsection{The $\Delta$LFBE feature}
The logarithmic Mel-filterbank energy (LFBE)  feature  can be computed as:
\vspace{-0.5em}
\begin{equation}
\vspace{-0.1em}
\label{eq:lfbe}
LFBE_i(x) = \log m_i^T \|\mathcal{F} (w \odot x)\|^2\mathrm{,}
\end{equation}
where $LFBE_i(x) $ denotes the $i$th LFBE band of a signal frame $x$, $m_i$ denotes the $i$th Mel filter, $\mathcal{F}$ denotes the Fourier transform, $w \odot x$ denotes an analysis window function $w$ element-wise multiplied by the signal $x$, and the squared modulus is applied element-wise.  

Assuming that the changes in the AFE gain scaled the volume level of the audio signal by a multiplicative constant $c$, then
\vspace{-0.5em}
\begin{equation}
\label{eq:global_c}
\begin{aligned}
LFBE_i(cx) &= \log m_i^T\|\mathcal{F} (w \odot cx)\|^2 \\
&=  \log m_i^T|c|^2\|\mathcal{F} (w \odot x)\|^2  \\
&=\log |c|^2+\log m_i^T\|\mathcal{F} (w \odot x)\|^2 \\
&=2\log|c| + LFBE_i(x)\mathrm{.}
\end{aligned}
\end{equation}
Therefore changing the volume of audio corresponds to adding a constant to all of the LFBE frames. 

The $\Delta$LFBE computes the pairwise difference between two contiguous LFBE frames, i.e. $
\Delta LFBE(x) = LFBE(x(t)) - LFBE(x(t-1))$. Therefore the constant get canceled, i.e. $\Delta LFBE(cx) = \Delta LFBE(x)$. 

In addition,
if a filter $h$ satisfies $\lvert \mathcal{F}(h)\rvert \approx c_i $ over the $i$th Mel band, then $LFBE_i(h *x)\approx 2\log\vert c_i\rvert + LFBE_i(x)$. This means the per-Mel-band approximate scaling factor $c_i$ will get canceled by the difference. Therefore the $\Delta$LFBE is approximately invariant to filters whose magnitude response varies slowly within each Mel band.

\subsection{Implementation}
Rather than modifying the feature extraction module to perform the delta operation, we propose two approaches to implemented it in a neural network (NN). The delta operation can either be implemented explicitly as a fixed layer, or implicitly by folding it into the first layer as a constraint. Both approaches apply to architectures such as CNN or fully-connected DNN.

For conventional 2D inputs , the delta operation is equivalent to convolving the input with a filter of size $2\times 1 \times 1$, where the dimensions correspond to time, frequency and channel. The filter weights are $[-1, 1]$, which means taking the difference along the time dimension. In training, this layer can be set as frozen during training to preserve the temporal delta operation.

For flattened inputs, applying the temporal delta operation is a sparse linear transformation which can be folded into the first layer. Let the receptive field contain $n$ frames of LFBE feature. Considering a vector composed of the $i$th LFBE-band coefficients from all the frames in the context window, i.e. $\mathbf{x}^i = [x_1^i, x_2^i, ..., x_n^i]^T$, taking the temporal delta can be viewed as multiplying a sparse matrix $\bar{D}$: 
\vspace{-0.5em}
\begin{equation}
\vspace{-0.5em}
\bar{D}\mathbf{x}^i=\begin{bmatrix}
-1&1&&&\\
&-1&1&&\\
 &&\cdots &\cdots&\\
 &&&-1&1
\end{bmatrix}_{(n-1) \times n} 
\begin{bmatrix}
x_1^i\\
x_2^i\\
\vdots\\
x_n^i
\end{bmatrix}_{\textstyle .} \quad
\end{equation}
For the LFBE coefficients of all of the $L$ bands, the temporal delta transformation can be written as 
\vspace{-0.5em}
\begin{equation}
\label{eq:delta}
\vspace{-0.5em}
\mathbf{D} = \begin{bmatrix}
\bar{D} &&&\\
&\bar{D} &&\\
&&...&\\
&&&\bar{D}
\end{bmatrix}_{(n-1)L\times nL\mathrm{.}}
\end{equation}
 
Let $\mathbf{W}_{R\times (n-1)L}$ denote the first  layer weight matrix where $R$ is the number of nodes in the first hidden layer. Let $\mathbf{V}_{R\times nL}=\mathbf{WD}$ be the equivalent first layer weight matrix combined with the temporal delta operation. Considering the $r$th row of $\mathbf{W}$, which can be written as $\mathbf{w}_r=[\mathbf{w}_r^1, \mathbf{w}_r^2, ..., \mathbf{w}_r^{L}]$, where $\mathbf{w}_r^{i} = [w_{r,1}^i, w_{r,2}^i, ..., w_{r, n-1}^i], i=1,2,...,L$, then
\vspace{-0.4em}
\begin{equation}
\vspace{-0.1em}
\begin{aligned}
\mathbf{w}_r\mathbf{D} =& \left[ \mathbf{w}_r^1\bar{D}, \mathbf{w}_r^2\bar{D}, ... ,\mathbf{w}_r^i\bar{D}, ...,\mathbf{w}_r^L\bar{D}  \right] \\
=&\left[ ...,-w_{r,1}^i, w_{r,1}^i-w_{r,2}^i, ..., w_{r,n-2}^i-w_{r,n-1}^i, w_{r,n-1}^i, ... \right]_{.}
\end{aligned}
\end{equation}
Notice that for each Mel band $i$, the sum is 
\vspace{-0.5em}
\begin{equation}
\label{eq:zero-sum}
\vspace{-0.1em}
-w_{r1}^i + w_{r1}^i-w_{r2}^i + ... +w_{r,n-2}^i-w_{r,n-1}^i+ w_{r,n-1}^i = 0.
\end{equation}
This means the temporal delta operation requires the per-Mel-band (row-block-wise) zero-sum constraint on rows of the first layer weight matrix $\mathbf{V}$. Conversely, a matrix $\mathbf{V}$ satisfying \eqref{eq:zero-sum} can be factored into $\mathbf{WD}$ where $\mathbf{D}$ is in the form of \eqref{eq:delta}. Therefore $\Delta$LFBE for flattened input can be implemented by adding a kernel constraint of per-Mel-band mean-subtraction on the rows of the first layer weight matrix $\mathbf{V}$.
\subsection{Relation to other work}
The spectral delta features are commonly used as supporting features 
such as  in MFCC for speech recognition tasks, as well as the only input for RNN-based musical onset detection  tasks \cite{Eyben2010}. The $\Delta$LFBE is a type of first-order spectral delta feature applied to LFBE. Our implementation simply embeds the delta operation into the computational graph of a neural network, therefore it is compatible with existing WW systems.

The 2D $\Delta$LFBE can be viewed as a special case of zero-mean convolution \cite{schluter2018zero} with per-Mel-band zero-mean constraints. In the general zero-mean convolution, since the filter covers a local time-frequency tile, it can cancel the constant gain change as in \eqref{eq:global_c}, however it will not cancel the per-Mel-band scaling changes. On the contrary, the  $\Delta$LFBE is invariant to the per-Mel-band scaling.

%% file: experiments.tex
\label{sec:expr}
\subsection{Model Architectures}
\begin{figure}[h]
\vspace{-0.75em}
\centering
\includegraphics[width=0.48\textwidth]{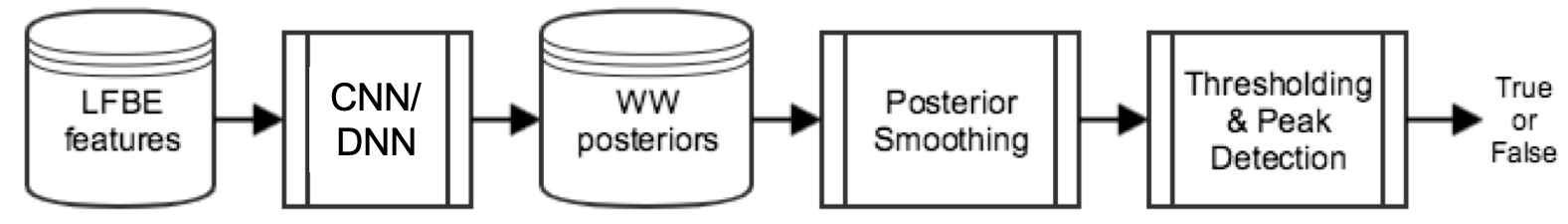}
\caption{The WW Spotter}
\label{fig:spotter}
\vspace{-0.75em}
\end{figure}

The wake word spotting system in this paper is composed of three main components: neural network, posterior smoothing and peak detection, as shown in Figure \ref{fig:spotter}. We experiment with two types of WW model architectures: fully-connected DNN and CNN, represented for low- and mid- computational budgets.  Table \ref{tab:architecture} summarizes the two model architectures for WW spotting, where ``f" denotes the fully-connected layer and ``c" denotes the convolution layer. 
\begin{table}[h]
\vspace{-0.75em}
\caption{Summary of WW Model Architectures}
\label{tab:architecture}
\vspace{-0.5em}
\begin{tabular}{cccccc}
Type&Layers&\# Par&\# Mul&Input&Smoothing \\
\hline
DNN&6f&222K&222K&540&WMA\\
CNN&5c3f&5.6M&6.7M&64x100&EMA\\
\end{tabular}
\vspace{-0.5em}
\end{table}

For decoding, the posteriors corresponding to the WW are first smoothed by a windowed moving average (WMA) or exponential moving average (EMA), then a thresholding with peak-detection algorithm is applied to infer the WW hypothesis.

The input audio to the WW spotter is re-sampled at 16KHz and represented by 16-bits signed integers. The LFBE feature is computed every 10 msec with a 25 msec analysis window denoted by $w$ in \eqref{eq:lfbe}. The input to the DNN is a 540-dimensional vector from a stack of 80 frames of 20-bands LFBE feature down-sampled by 3 and the input to the CNN model is 100 frames of 64-bands LFBE feature.  

\begin{figure}[t]
\vspace{-1em}
\centering
\includegraphics[width=0.45\textwidth]{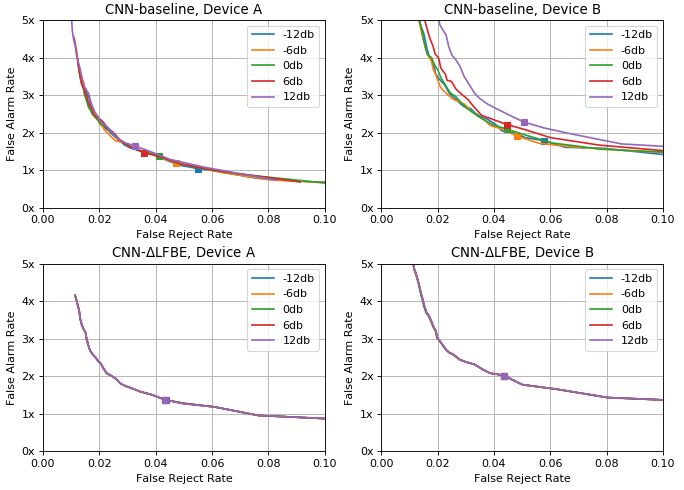}
\caption{Spotting ``Alexa" with CNN WW models. In both the bottom figures (using $\Delta$LFBE), the DET curves are on top of one another, indicating that the WW model with $\Delta$LFBE is robust to AFE changes.}
\label{fig:cnn}
\vspace{-1.5em}
\end{figure}
\vspace{-0.5em}
\subsection{Experimental Setup}
The data set for training the WW models is composed of 8.7M
annotated speech utterances with 10\% augmented with noise corruption to improve the general noise robustness in far-field. For model training we use the standard cross-entropy loss and the ADAM optimizer with learning rate of 0.001 and dropout probability of 0.3. Batch-normalization is applied to all hidden layers.

The original test data is collected from two types of devices: Type A  and Type B such that the training set contains data from A but not from B. We used device Type A to study the effects of AFE changes for an existing WW model; and used Type B to study the case when deploying a WW model to a new device. The original test set contains 24341 streams from A and 14495 streams from B. The evaluation metrics reported here are False Rejection Rate (FRR) and false alarm rate (FAR). Absolute values of FAR in this paper are anonymized for confidentiality reasons. The range 
of used FAR corresponds to the range normally used for production
keyword spotting models. We report the two metrics in a detection-error-tradeoff (DET) curve as we tune the operating point (OP) for each model. \footnote{The experimental results in this paper do not represent the performance of the production Alexa system.} 

\vspace{-0.5em}
\subsection{Data simulation}
To create our evaluation data sets representing different AFE gains, we simulate the effects of gain changes by varying the power level of the audio data from the original test set with amplification or attenuation by certain decibels (dB).  For 16-bits signed-integer PCM signals, the full-scale power is $10\log_{10}(|2^{15}|^2)$ $\approx  90$dB. We took the scaling step size equivalent to 1 bit shift, which corresponds to double or half the volume in magnitude,  and is approximately 6dB ($10\log_{10} 2^2$) in power. We sweep the gain changes in \{-12dB, -6dB, 0dB, 6dB, 12dB\} and created 5 versions of test sets from the original test set collected from device types A and B, respectively.

The scaling in volume may introduce clipping (in amplification) and quantization (in attenuation) noise. We decouple the clipping and quantization with gain changes by applying hard dynamic range compression (HDRC) to the audio before changing the gain. That is, we first shrink the range of the signal to only occupy bits from $b$th to $(B-b)$th bits by zeroing out the least and most significant $b$ bits, where $b$=2 corresponds to maximum bit shifts in our experiments, and $B$=15 is the maximum number of bits of a 16-bits signed-integer representation. Note that only zeroing out the leaset significant bits (LSBs) is equivalent to quantize the signal into $(B-b)$ bits, and only zeroing out the most significant bits (MSBs) is equivalent to clipping the signal at $(B-b)$ bits.

\begin{figure}[t]
\vspace{-1em}
\centering
\includegraphics[width=0.45\textwidth]{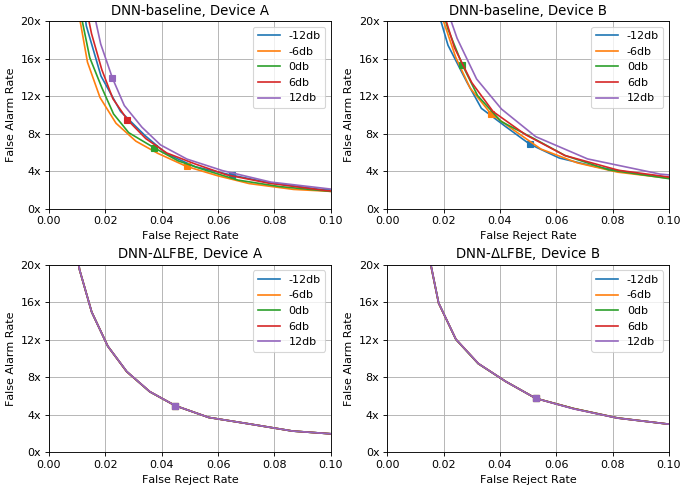}
\caption{Spotting ``Alexa" with DNN WW models. In both the bottom figures (using $\Delta$LFBE), the DET curves are right on top of one another, suggesting the performance is preserved regardless of AFE gain changes.}
\label{fig:dnn}
\vspace{-1.5em}
\end{figure}
\vspace{-0.5em}
\subsection{Wake word spotting results}
We run experiments of spotting the WW ``Alexa" against the test data with simulated gain changes in \{-12dB, -6dB, 0dB, 6dB, 12dB\}. Figures \ref{fig:cnn} and \ref{fig:dnn} show results with the baseline model and $\Delta$LFBE model of CNN and DNN architectures, respectively. In both figures, the top row corresponds to the baseline model and the bottom row corresponds to the $\Delta$LFBE model; the left column corresponds to device type A, which the models are trained on, and the right column corresponds to device type B which is unseen when training the WW model. The OP is denoted by the square-shaped dots on the DET curves.

It can be seen that in both the CNN and DNN cases, the baseline models are sensitive to the gain changes. For CNN-baseline on device A (B), the OP shifts caused relative FAR increase by 19.0\% (9.0\%) or FRR increase by 34.3\% (29.5\%)when the gain level varies up to $\pm$12dB. Choosing a new OP along the curve restores the performance, but requires extra tuning effort. For the unknown device B, the gain changes not only caused OP shift, but also the DET curves, which means it is not possible to restore the performance with OP tuning. 

In contrast, the  models using $\Delta$LFBE show invariance to the gain changes. The DET curves and OPs do not move w.r.t the gain changes,  indicating that the desired performance is preserved. The results verify the theoretical analysis in Section \ref{sec:dlfbe}.

It is worth noting that for an unknown device such as device B, models with $\Delta$LFBE perform slightly better than the baseline models by comparing the curves at 0dB. This suggests that $\Delta$LFBE not only is invariant to AFE gain changes, but also generalizes well to new devices.

\begin{figure}[ht]
\vspace{-1em}
\centering
\includegraphics[width=0.4\textwidth]{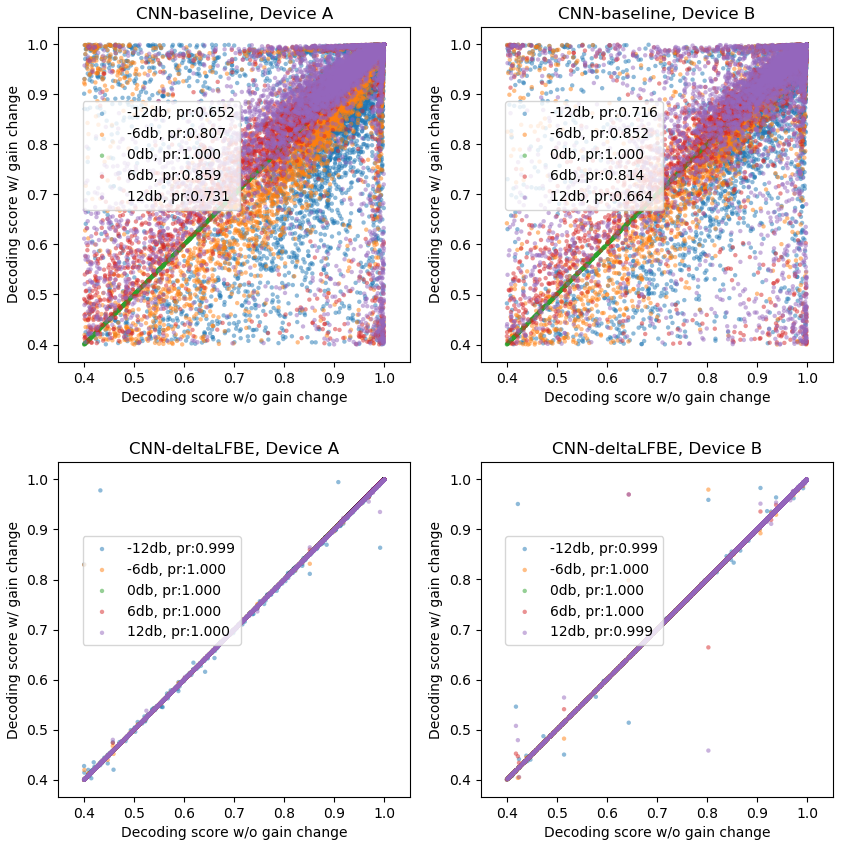}
\caption{Scatter plots of decoding scores from CNN.}
\label{fig:scatter}
\vspace{-2em}
\end{figure}
We further analyze the decoding scores with and without the gain changes. The decoding score is the value of the smoothed posteriors picked by the peak detection algorithm.  Figure \ref{fig:scatter} shows the scatter plots of the decoding scores with gain changes against the scores without gain changes.  It can be seen that the decoding scores from the $\Delta$LFBE models are strongly correlated as the scatters are concentrated on the diagonal line. The Pearson correlation coefficients (denoted by ``pr" in Figure \ref{fig:scatter}) also suggest that decoding scores of the $\Delta$LFBE model are less sensitive to the gain changes. 

\vspace{-0.5em}
\subsection{Effects of data simulation}
\begin{figure}[ht]
\vspace{-1.0em}
\centering
\includegraphics[width=0.45\textwidth]{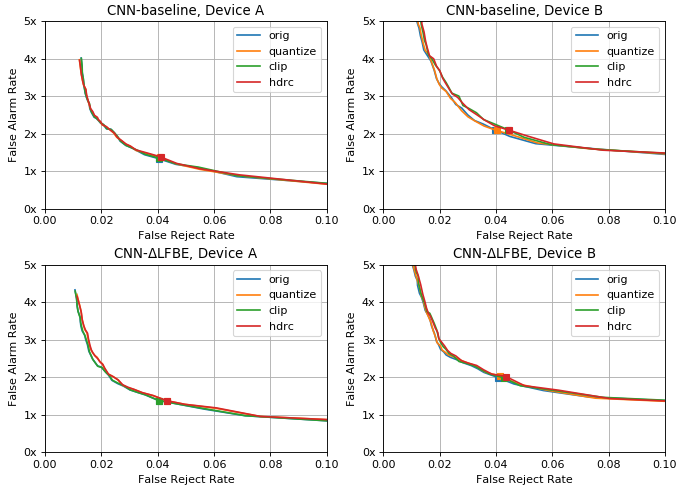}
\caption{Effects of quantization, clipping and hard dynamic range compression (hdrc).}
\label{fig:hdrc}
\vspace{-1.5em}
\end{figure}

We are also interested in how realistic it is for the simulated gain change in real data in terms of WW spotting performance. Figure \ref{fig:hdrc} show the DET curves of the original, quantized, clipped and HDRC-processed.  We can see that the gaps between the DET curves for both models are negligible, indicating the quantization and clipping noise have limited affects on WW spotting on these data sets, therefore the simulated data is reliable to mimic the AFE gain changes in reality. 
\vspace{-0.5em}

Moreover, comparing the two plots in the right column suggests that $\Delta$LFBE helps to close the small gap between the curves of quantized and HDRC, indicating the $\Delta$LFBE further removes the disparity in data simulation.

\vspace{-0.5em}
\subsection{Public dataset: audio tagging on DCASE2017}
\begin{figure}[tb]
\vspace{-1.0em}
\centering
\includegraphics[width=0.4\textwidth]{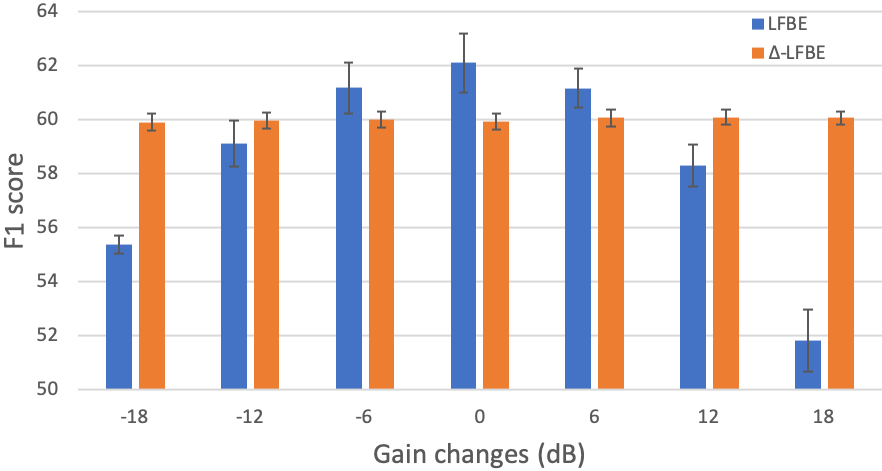}
\caption{Audio tagging results on DCASE2017 Task 2A.}
\label{fig:aed}
\vspace{-1.5em}
\end{figure} 
\vspace{-0.5em}
We also study the generalizability of the $\Delta$LFBE feature on an audio tagging task using a public dataset: DCASE2017 Task 2A \cite{DCASE2017challenge}.
It contains 17 classes of warning and vehicle sounds related to driving environments for smart car applications.
Classification F1 score is reported. The model is a ``ResNet" in ~\cite{DenseNetAED}  which is similar to ResNet-18 in~\cite{ResNet} with fewer number of filters in each block (from [64, 128, 256, 512] to [28, 56, 112, 224]). 
We train models using ADAM optimizer with the initial learning rate of 0.01. 
For each type of features we ran five trials to compensate the randomness in training process.
For evaluation, on top of the original test set (0dB), we simulated test sets with gain changes in \{-18dB, -12dB, -6dB, 6dB, 12dB, 18dB\}.

The bars and error bars shown in Figure \ref{fig:aed} are the averaged F1 and standard deviation over 5 trials.
Models using LFBE features (blue bars) are very sensitive to gain changes. 
There are significant relative decreases in F1 score for gain of 18dB (-10.27\%) and -18dB (-6.72\%).
On the contrary, models using $\Delta$LFBE (orange bars) have consistent performance across different gains.
Shorter error bars for models using $\Delta$LFBE indicate that it brings less variance across different trials.

One interesting observation is that when the gain is small (-6dB, 0dB, 6dB), models using LFBE outperform $\Delta$LFBE. We hypothesize that this performance degradation is due to the lack of proper finetuning on hyper-parameters.
This proof of concept experiment is designed to show the robustness to the effects of AFE gain changes on a public datasets, so we did not put much efforts to finetune the hyper-parameters.

%% file: 1992_camera_ready.bbl
\begin{thebibliography}{10}
\providecommand{\url}[1]{#1}
\csname url@samestyle\endcsname
\providecommand{\newblock}{\relax}
\providecommand{\bibinfo}[2]{#2}
\providecommand{\BIBentrySTDinterwordspacing}{\spaceskip=0pt\relax}
\providecommand{\BIBentryALTinterwordstretchfactor}{4}
\providecommand{\BIBentryALTinterwordspacing}{\spaceskip=\fontdimen2\font plus
\BIBentryALTinterwordstretchfactor\fontdimen3\font minus
  \fontdimen4\font\relax}
\providecommand{\BIBforeignlanguage}[2]{{%
\expandafter\ifx\csname l@#1\endcsname\relax
\typeout{** WARNING: IEEEtran.bst: No hyphenation pattern has been}%
\typeout{** loaded for the language `#1'. Using the pattern for}%
\typeout{** the default language instead.}%
\else
\language=\csname l@#1\endcsname
\fi
#2}}
\providecommand{\BIBdecl}{\relax}
\BIBdecl

\bibitem{Chen2014_DNN}
G.~Chen, C.~Parada, and G.~Heigold, ``Small-footprint keyword spotting using
  deep neural networks,'' in \emph{IEEE ICASSP}, 2014.

\bibitem{Sainath2015_CNN}
T.~N. Sainath and C.~Parada, ``Convolutional neural networks for
  small-footprint keyword spotting,'' in \emph{INTERSPEECH}, 2015.

\bibitem{He2017_Seq2Seq}
Y.~He, R.~Prabhavalkar, K.~Rao, W.~Li, A.~Bakhtin, and I.~McGraw, ``Streaming
  small-footprint keyword spotting using sequence-to-sequence models,'' in
  \emph{ASRU}, 2017.

\bibitem{Sun2017_TDNN}
M.~Sun, D.~Snyder, Y.~Gao, V.~K. Nagaraja, M.~Rodehorst, S.~Panchapagesan,
  N.~Strom, S.~Matsoukas, and S.~Vitaladevuni, ``Compressed time delay neural
  network for small-footprint keyword spotting,'' in \emph{INTERSPEECH}, 2017.

\bibitem{Kumatani2017_RAW}
K.~Kumatani, S.~Panchapagesan, M.~Wu, M.~Kim, N.~Strom, G.~Tiwari, and
  A.~Mandai, ``Direct modeling of raw audio with dnns for wake word
  detection,'' in \emph{ASRU}, 2017, pp. 252--257.

\bibitem{Shan2018_Attn}
C.~Shan, J.~Zhang, Y.~Wang, and L.~Xie, ``Attention-based end-to-end models for
  small-footprint keyword spotting,'' in \emph{INTERSPEECH}, 2018.

\bibitem{Guo2018_Highway}
J.~Guo, K.~Kumatani, M.~Sun, M.~Wu, A.~Raju, N.~Strom, and A.~Mandal,
  ``Time-delayed bottleneck highway networks using a dft feature for keyword
  spotting,'' in \emph{IEEE ICASSP}, 2018.

\bibitem{Wu2018_Monophone}
M.~Wu, S.~Panchapagesan, M.~Sun, J.~Gu, R.~Thomas, S.~Vitaladevuni,
  B.~Hoffmeister, and A.~Mandal, ``Monophone-based background modeling for
  two-stage on-device wake word detection,'' in \emph{IEEE ICASSP}, 2018.

\bibitem{Panchi2016_Multitask}
S.~Panchapagesan, M.~Sun, A.~Khare, S.~Matsoukas, A.~Mandal, B.~Hoffmeister,
  and S.~Vitaladevuni, ``Multi-task learning and weighted cross-entropy for
  dnn-based keyword spotting,'' in \emph{INTERSPEECH}, 2016.

\bibitem{Tucker2016_SVD}
G.~Tucker, M.~Wu, M.~Sun, S.~Panchapagesan, G.~Fu, and S.~Vitaladevuni, ``Model
  compression applied to small-footprint keyword spotting,'' in
  \emph{INTERSPEECH}, 2016.

\bibitem{Raju2018_Augm}
A.~Raju, S.~Panchapagesan, X.~Liu, A.~Mandal, and N.~Strom, ``Data augmentation
  for robust keyword spotting under playback interference,'' in
  \emph{arXiv:1808.00563}, 2018.

\bibitem{Li2018_TS}
J.~Li, R.~Zhao, Z.~Chen, C.~Liu, X.~Xiao, G.~Ye, and Y.~Gong, ``Developing
  far-field speaker system via teacher-student learning,'' in
  \emph{INTERSPEECH}, 2018.

\bibitem{Chen2018_MAML}
Y.~Chen, T.~Ko, L.~Shang, X.~Chen, X.~Jiang, and Q.~Li, ``Meta learning for
  few-shot keyword spotting,'' in \emph{arXiv preprint arXiv:1812.10233}, 2018.

\bibitem{Chhetri2018}
A.~{Chhetri}, P.~{Hilmes}, T.~{Kristjansson}, W.~{Chu}, M.~{Mansour}, X.~{Li},
  and X.~{Zhang}, ``Multichannel audio front-end for far-field automatic speech
  recognition,'' in \emph{2018 26th European Signal Processing Conference
  (EUSIPCO)}, 2018, pp. 1527--1531.

\bibitem{perez2011automatic}
J.~P.~A. P{\'e}rez, S.~C. Pueyo, and B.~C. L{\'o}pez, \emph{Automatic gain
  control}.\hskip 1em plus 0.5em minus 0.4em\relax Springer, 2011.

\bibitem{prabhavalkar2015automatic}
R.~Prabhavalkar, R.~Alvarez, C.~Parada, P.~Nakkiran, and T.~N. Sainath,
  ``Automatic gain control and multi-style training for robust small-footprint
  keyword spotting with deep neural networks,'' in \emph{2015 IEEE
  International Conference on Acoustics, Speech and Signal Processing
  (ICASSP)}.\hskip 1em plus 0.5em minus 0.4em\relax IEEE, 2015, pp. 4704--4708.

\bibitem{wang2017trainable}
Y.~Wang, P.~Getreuer, T.~Hughes, R.~F. Lyon, and R.~A. Saurous, ``Trainable
  frontend for robust and far-field keyword spotting,'' in \emph{2017 IEEE
  International Conference on Acoustics, Speech and Signal Processing
  (ICASSP)}.\hskip 1em plus 0.5em minus 0.4em\relax IEEE, 2017, pp. 5670--5674.

\bibitem{battenberg2017reducing}
E.~Battenberg, R.~Child, A.~Coates, C.~Fougner, Y.~Gaur, J.~Huang, H.~Jun,
  A.~Kannan, M.~Kliegl, A.~Kumar \emph{et~al.}, ``Reducing bias in production
  speech models,'' \emph{arXiv preprint arXiv:1705.04400}, 2017.

\bibitem{lostanlen2018per}
V.~Lostanlen, J.~Salamon, M.~Cartwright, B.~McFee, A.~Farnsworth, S.~Kelling,
  and J.~P. Bello, ``Per-channel energy normalization: Why and how,''
  \emph{IEEE Signal Processing Letters}, vol.~26, no.~1, pp. 39--43, 2018.

\bibitem{lostanlen2019long}
V.~Lostanlen, K.~Palmer, E.~Knight, C.~Clark, H.~Klinck, A.~Farnsworth,
  T.~Wong, J.~Cramer, and J.~P. Bello, ``Long-distance detection of bioacoustic
  events with per-channel energy normalization,'' in \emph{Acoustic Scenes and
  Events 2019 Workshop (DCASE2019)}, 2019, p. 144.

\bibitem{lehner2018online}
B.~Lehner, J.~Schl{\"u}ter, and G.~Widmer, ``Online, loudness-invariant vocal
  detection in mixed music signals,'' \emph{IEEE/ACM Transactions on Audio,
  Speech, and Language Processing}, vol.~26, no.~8, pp. 1369--1380, 2018.

\bibitem{Atal1974cmn}
\BIBentryALTinterwordspacing
B.~Atal, ``Effectiveness of linear prediction characteristics of the speech
  wave for automatic speaker identification and verification,'' \emph{The
  Journal of the Acoustical Society of America}, vol.~55, no.~6, p.
  1304—1322, June 1974. [Online]. Available:
  \url{https://doi.org/10.1121/1.1914702}
\BIBentrySTDinterwordspacing

\bibitem{liu1993cmn}
\BIBentryALTinterwordspacing
F.-H. Liu, R.~M. Stern, X.~Huang, and A.~Acero, ``Efficient cepstral
  normalization for robust speech recognition,'' in \emph{{H}uman {L}anguage
  {T}echnology: Proceedings of a Workshop Held at Plainsboro, New Jersey, March
  21-24, 1993}, 1993. [Online]. Available:
  \url{https://www.aclweb.org/anthology/H93-1014}
\BIBentrySTDinterwordspacing

\bibitem{Rehr2015CNS}
R.~{Rehr} and T.~{Gerkmann}, ``Cepstral noise subtraction for robust automatic
  speech recognition,'' in \emph{2015 IEEE International Conference on
  Acoustics, Speech and Signal Processing (ICASSP)}, 2015, pp. 375--378.

\bibitem{kalinli2019parametric}
O.~Kalinli, G.~Bhattacharya, and C.~Weng, ``Parametric cepstral mean
  normalization for robust speech recognition,'' in \emph{ICASSP 2019-2019 IEEE
  International Conference on Acoustics, Speech and Signal Processing
  (ICASSP)}.\hskip 1em plus 0.5em minus 0.4em\relax IEEE, 2019, pp. 6735--6739.

\bibitem{cascade2017}
\BIBentryALTinterwordspacing
A.~Gruenstein, R.~Alvarez, C.~Thornton, and M.~Ghodrat, ``A cascade
  architecture for keyword spotting on mobile devices,'' 2017. [Online].
  Available: \url{https://arxiv.org/abs/1712.03603}
\BIBentrySTDinterwordspacing

\bibitem{Eyben2010}
F.~Eyben, S.~Bock, B.~Schuller, , and A.~Graves, ``Universal onset detection
  with bidirectional long short-term memory neural networks,'' in
  \emph{International Society for Music Information Retrieval Conference
  (ISMIR)}, 2010, p. 589–594.

\bibitem{schluter2018zero}
J.~Schl{\"u}ter and B.~Lehner, ``Zero-mean convolutions for level-invariant
  singing voice detection.'' in \emph{ISMIR}, 2018, pp. 321--326.

\bibitem{DCASE2017challenge}
A.~Mesaros, T.~Heittola, A.~Diment, B.~Elizalde, A.~Shah, E.~Vincent, B.~Raj,
  and T.~Virtanen, ``{DCASE} 2017 challenge setup: Tasks, datasets and baseline
  system,'' in \emph{DCASE}, 2017, pp. 85--92.

\bibitem{DenseNetAED}
C.~Kao, B.~Shi, M.~Sun, and C.~Wang, ``A joint framework for audio tagging and
  weakly supervised acoustic event detection using densenet with global average
  pooling,'' in \emph{INTERSPEECH}, 2020.

\bibitem{ResNet}
K.~He, X.~Zhang, S.~Ren, and J.~Sun, ``Deep residual learning for image
  recognition,'' in \emph{IEEE CVPR}, 2016, pp. 770--778.

\end{thebibliography}
